\documentclass[showpacs,amssymb,twocolumn]{revtex4}

\usepackage{graphicx}
\usepackage{dcolumn}
\usepackage{bm}
\usepackage{amsmath}

\begin{document}

\def\aj{AJ}%
\def\actaa{Acta Astron.}%
\def\araa{ARA\&A}%
\def\apj{ApJ}%
\def\apjl{ApJ}%
\def\apjs{ApJS}%
\def\ao{Appl.~Opt.}%
\def\apss{Ap\&SS}%
\def\aap{A\&A}%
\def\aapr{A\&A~Rev.}%
\def\aaps{A\&AS}%
\def\azh{AZh}%
\def\baas{BAAS}%
\def\bac{Bull. astr. Inst. Czechosl.}%
\def\caa{Chinese Astron. Astrophys.}%
\def\cjaa{Chinese J. Astron. Astrophys.}%
\def\icarus{Icarus}%
\def\jcap{J. Cosmology Astropart. Phys.}%
\def\jrasc{JRASC}%
\def\mnras{MNRAS}%
\def\memras{MmRAS}%
\def\na{New A}%
\def\nar{New A Rev.}%
\def\pasa{PASA}%
\def\pra{Phys.~Rev.~A}%
\def\prb{Phys.~Rev.~B}%
\def\prc{Phys.~Rev.~C}%
\def\prd{Phys.~Rev.~D}%
\def\pre{Phys.~Rev.~E}%
\def\prl{Phys.~Rev.~Lett.}%
\def\pasp{PASP}%
\def\pasj{PASJ}%
\def\qjras{QJRAS}%
\def\rmxaa{Rev. Mexicana Astron. Astrofis.}%
\def\skytel{S\&T}%
\def\solphys{Sol.~Phys.}%
\def\sovast{Soviet~Ast.}%
\def\ssr{Space~Sci.~Rev.}%
\def\zap{ZAp}%
\def\nat{Nature}%
\def\iaucirc{IAU~Circ.}%
\def\aplett{Astrophys.~Lett.}%
\def\apspr{Astrophys.~Space~Phys.~Res.}%
\def\bain{Bull.~Astron.~Inst.~Netherlands}%
\def\fcp{Fund.~Cosmic~Phys.}%
\def\gca{Geochim.~Cosmochim.~Acta}%
\def\grl{Geophys.~Res.~Lett.}%
\def\jcp{J.~Chem.~Phys.}%
\def\jgr{J.~Geophys.~Res.}%
\def\jqsrt{J.~Quant.~Spec.~Radiat.~Transf.}%
\def\memsai{Mem.~Soc.~Astron.~Italiana}%
\def\nphysa{Nucl.~Phys.~A}%
\def\physrep{Phys.~Rep.}%
\def\physscr{Phys.~Scr}%
\def\planss{Planet.~Space~Sci.}%
\def\procspie{Proc.~SPIE}%
\let\astap=\aap
\let\apjlett=\apjl
\let\apjsupp=\apjs
\let\applopt=\ao

\title{Testing the Newton second law in the regime of small accelerations}
\author{V. A. De Lorenci}
 \email{delorenci@unifei.edu.br}
\affiliation{Institute of Science, Federal University of Itajub\'a, 
37500-903 Itajub\'a, M. G., Brazil,}
\affiliation{PH Department, TH Unit, CERN, 1211 Geneva 23, Switzerland}
\author{M. Fa\'{u}ndez-Abans}
 \email{max@lna.br}
 \affiliation{Laborat\'orio Nacional de Astrof\'{\i}sica, Rua Estados Unidos 154, bairro das
Na\c{c}\~oes, 37504-364 Itajub\'a, MG, Brazil}
\author{J. P. Pereira}
 \email{jonas.pereira@yahoo.com.br}
\affiliation{Institute of Science, Federal University of Itajub\'a, 
37500-903 Itajub\'a, M. G., Brazil,}

\date{\today}

\begin{abstract}
It has been pointed out that the Newtonian second law can be tested in the very small acceleration regime by using the combined movement of the Earth and Sun around the Galactic center of mass. It has been shown that there are only two brief intervals during the year in which the experiment can be completed, which correspond to only two specific spots on the Earth surface. An alternative experimental setup is presented to allow the measurement to be made on Earth at any location and at any time.
\end{abstract}

\maketitle


\section{Introduction}
\label{I}

As is observationally known (Rubin \cite{rubin1983a,rubin1983b}; Cowsik \& Ghosh \cite{cg1986}; Famaey et al. \cite{fam2007}; and Gentile et al. \cite{gentile2007}), the rotation curve of galactic mass distribution cannot be interpreted in terms of  Newtonian classical dynamics. Far mass seems to exist  than directly inferred for luminous mass, which consists of stars, planets, dust, gas, and molecular clouds. To solve this theoretically unexpected result two possible scenarios have been examined in the literature. The first, which has been widely accepted, consists of postulating the existence of a fundamentally different kind of matter: the so-called dark matter. This type of matter is expected to interact with ordinary matter only gravitationally. Thus, it should not be directly detected by modern telescopes since it should not emit any radiation. Experiments continue to be performed to detect dark matter. One possible explanation of dark matter involves an extension of the standard model of particle physics to include weakly interacting massive particles (WIMPs). If dark matter exists, it is expected to be produced by the LHC experiment at CERN, which should take place imminently, or inside future high-energy colliders (Baltz et al. \cite{baltz2006}). 
Accretion of dark matter onto a central black hole at the Galactic center may lead a density spike in the dark matter distribution, which would result in a high dark matter annihilation rate with observational consequences (Ullio et al. \cite{ullio2001}). 
In the context of cold dark matter (CDM) cosmological models, structures are generated hierarchically and local dark matter distribution depends on the fate of the first WIMP micro-halos to form. The evolution of these micro-halos is being considered and may hold the key to dark matter detections (Angus \& Zhao \cite{angus2007b}; Green \cite{green2007}, and references therein).
Results based on the weak lensing observations of the so-called Bullet Cluster (1E0657-588, z=0.296) were claimed to represent empirical proof of the existence of dark matter (Clowe et al. \cite{clowe2006}). However, a possible conflicting behavior measured in a similar merging system was pointed out in Mahdavi et al. (\cite{mahdavi2007}). 
Penny et al. (\cite{penny09}), analyzed observations by the high-resolution Hubble Space Telescope Advanced Camera for Surveys, that is WFS imaging in the F555W and F814W bands of five fields about the core of the Perseus cluster of galaxies, and proposed that a large proportion of dark matter could explain the absence of tidal disruption in the core dwarf galaxies.   

The second possible scenario for explaining the apparent mass discrepancies inside galaxies consists of a modification to the Newtonian dynamics in the limit of low accelerations (Milgrom \cite{milgrom1983a,milgrom1983b}). The main assumption of this proposed \emph{Modified Newtonian Dynamics} (MOND) is that the well-established classical second law of dynamics would change to the form
%
\begin{equation}
\vec{F} = m \mu(a/a_o) \vec{a},
\label{mond2ndlaw}
\end{equation}

\noindent where $a_o$ is a new physical constant with the 
dimension of acceleration and the function $\mu$ satisfies the 
phenomenological conditions
%
\begin{equation}
\mu(a/a_o) =
\left\{
     \begin{array}{cc}
                      1,     &    a >> a_o \\
                      a/a_o, &    a << a_o.  
     \end{array}
\right.
\label{matrix}
\end{equation}

Following Milgrom (\cite{milgrom1983a,milgrom1983b}) and Bekenstein \& Milgrom (\cite{bekenstein1984}), two possible approaches could be considered in this scenario. The first refers to a modification to the gravity sector that does not change Newton's second law, and would imply changing the gravitational force from $\vec{F} = m\vec{g}$ \footnote{Here $m$ represents the gravitational mass and $\vec{g} = -\nabla\phi$ is the conventional gravitational field, deduced from a mass distribution using Poisson's equation.} to $\vec{F}=m\vec{g'}$, where $\vec{g'}$  is related to $\vec{g}$ by $\mu(g'/a_0) \vec{g'} = \vec{g}$. The second possible approach refers to changing the dynamics itself, i.e., the law of inertia, without changing the gravitational force. Although these two approaches lead to equivalent conclusions in considering gravitational systems, they conceptually differ. The former involves a change specifically in the gravitational force, while the latter corresponds to a fundamental modification of the dynamics of any force involved.

Phenomenologically, MOND is able to fit rotation curve data of observed spiral galaxies based only on the observed baryonic mass distribution (see Sanders \& McGaugh \cite{sanders2002}, and references therein). Furthermore, it was shown that the results of the Bullet Cluster measurements are consistent with MOND dynamics of single clusters (Angus et al. \cite{angus2007}). Numerical simulations with MOND were produced by Tiret \& Combes (\cite{tc07}) by taking into account gas dissipation and star formation. Their results show that MOND implies longer merger times, which solve the problem of the apparent excess of compact groups of galaxies observed today. In addition, galaxies in MOND are found to form stronger bars, and more rapidly than in Newtonian dynamics with dark matter (Tiret \& Combes 2007). In a dissipationless evolution of a MOND-dominated region in an expanding Universe, Sanders (\cite{san08}) demonstrated that the final virialized objects resemble elliptical galaxies with clearly defined relationships between  mass, radius, and velocity dispersion. Zhao et al. (\cite{zhao2008}) studied the properties of galaxies formed in the Tensor Vector Scalar (TeVeS) framework (Bekenstein \cite{bekenstein2004}) with analytical modeling and hydrodynamical simulations, comparing them with those in a $\Lambda$CDM framework in the context of the formation of both high surface brightness bulges and elliptical galaxies. As a result, they found that the use of a MOND scalar field produces tight correlations in galaxy properties because in this model the effective halo is not as scattered and free as in CDM ones.

Since the possible corrections implied by this approach would be applicable only at very low accelerations ($a_o \approx 10^{-10}m/s^2$), it should be improbable to detect the effects of MOND in terrestrial-based experiments. Nevertheless, a possible way of measuring the tiny effects of the proposed modified dynamics on the motion of test particles in terrestrial laboratory experiments was presented by Ignatiev (\cite{ignatiev2007}). It was shown that the combination of the motion of the Sun around the galaxy with the motion of translation and rotation of the Earth produces two spots on the Earth, symmetric with respect to the latitude coordinates, in which the MOND effects could be probed. In these two spots, of around $0.1m^3$ occurring for only a few seconds in each equinox date, static particles would experience a subtle acceleration caused by the MOND regime. 

The main objective of this paper is to extend the ideas presented in Ignatiev (\cite{ignatiev2007}) and ensure that the above mentioned experiment is feasible any time on other places on Earth. It is expected that the results of this measurement would be valuable in deciding the validity of Newtonian dynamics even in the regime of very small accelerations. It should be emphasized that instead of trying to explain the unusual behavior of specific astronomical systems, this type of measurement should decide which fundamental laws of physics should be applied.

\section{Testing the Newton's second law in the small acceleration regime}

In the MOND scenario, appreciable corrections to  
classical dynamics are required only in the regime of small accelerations $a_o$. 
As proposed by Ignatiev (\cite{ignatiev2007}), using a reference 
frame ($S_o$) fixed on the center of mass of the Galaxy, it is possible 
to achieve extremely small total accelerations 
\footnote{
It should be emphasized that the existence of small accelerations with respect to the laboratory is an insufficient condition to achieve the MOND regime. It is also necessary to consider external accelerations (caused by objects outside the laboratory), which would explain some observed phenomena in light of MOND (Famaey et al. \cite{famaey2007}; Wu et al. \cite{wu2007,wu2008}).}
by considering the combined motion of the Sun and the Earth. As presented in Ignatiev (\cite{ignatiev2007}), the approximate expression for the acceleration of a test particle at rest on the Earth's surface relative to $S_o$ is given by
%
\begin{equation}
\vec{a}_{\scriptscriptstyle S_o} \approx \vec{a}_1(t) +  \vec{a}_2(t) + 
\vec{\omega}\times(\vec{\omega}\times\vec{r}) \doteq \vec{A},
\label{1}
\end{equation}
where $\vec{\omega}$ is the angular velocity of Earth's rotation, $\vec{r}$ the position of the particle with respect to Earth's center-of-mass, and $ \vec{a}_1(t)$ and $\vec{a}_2(t)$ represent the accelerations of the Earth with respect to the Sun and of the Sun with respect to $S_o$, respectively. The MOND regime is expected to occur when $\vec{a}_{\scriptscriptstyle S_o} \approx 0$. This condition can be verified only if $\vec{a}_1(t) +  \vec{a}_2(t)$ become almost orthogonal to the angular velocity $\vec{\omega}$. In this situation,  $\vec{a}_1(t) +  \vec{a}_2(t)$ will be anti-parallel to the inertial acceleration $\vec{\omega}\times(\vec{\omega}\times\vec{r})$. Considering the dynamic nature of the entire system, this condition will be applicable for only few instants. Thus, provided the magnitude of the involved quantities is such that ${a}_{\scriptscriptstyle S_o} << a_o$, the MOND regime would be achieved. It was indeed shown (Ignatiev \cite{ignatiev2007}) that this condition can be verified on Earth's surface for a time interval of the order of $1s$ on each equinox date in a small region located on the Earth with coordinates $\theta \approx \pm 79^o50'$ and $\varphi\approx 56^oW$ (2008 equinox). Furthermore, the MOND correction would be significant in a space box of volume of approximately $0.1m^3$.

\section{An alternative experimental setup}

To extend the possibilities of testing the Newtonian dynamics in the small acceleration regime, we include a new rotating object in the above experimental setup.  For simplicity, it is considered to be a ring of radius $R$ rotating about its axial axes with angular velocity $\vec{\Omega}$. The direction of the rotation axis can be chosen freely and fixed by assuming the ring to be part of a gyroscopic device.  If the ring rotates with constant velocity, a test particle at rest with respect to its surface, identified by the position $\vec{R}$ will experience, in addition to $\vec{A}$, an inertial acceleration 
$
\vec{A}'\doteq\vec{\Omega}\times(\vec{\Omega}\times\vec{R}) 
$. 
The  Coriolis contribution $2\vec{\omega}\times\vec{v}$ must appear in the
expression for $\vec{A}$, since the test particle will have non-zero motion with
respect to Earth's surface. Thus, with respect to the galactic center-of-mass, the resulting acceleration of the test particle will be given by 
%
%
\begin{equation}
\vec{a}_{\scriptscriptstyle S_o} \approx \vec{A} + \vec{A}'. 
\label{2}
\end{equation}
The vector $\vec{\Omega}$ can be adjusted to ensure that $a_{\scriptscriptstyle S_o}\approx 0$ in a small region on the ring surface, at any time and location on the Earth. This will be the case provided that $\vec{\Omega}$ and $R$ can take values such that
%
%
\begin{equation}
\vec{\Omega}\times(\vec{\Omega}\times\vec{R})
\approx \vec{a}_1(t) +  \vec{a}_2(t) + 
\vec{\omega}\times(\vec{\omega}\times\vec{r}) + 2\vec{\omega}\times\vec{v}.
\label{3}
\end{equation}
The acceleration $\vec{A}$ of the combined motion of Earth (rotation and translation) and Sun towards the galaxy center-of-mass can be quantified by using astrophysical data. Thus, the inertial acceleration added by the rotating ring can be used to meet the condition expressed by Eq. (\ref{3}). 

Some estimates show that in order to achieve the regime of small acceleration $a_o$, it is sufficient to produce an additional inertial acceleration $A'$ in the interval
%
\begin{equation}
5\times 10^{-3}m/s^2 \lesssim A' \lesssim 3\times 10^{-2}m/s^2,
\label{app}
\end{equation} 
where the limit inferior corresponds to a ground-based experiment about the Earth's poles and the limit superior relates to a ground-based experiment in the equatorial band. 

Now let us suppose that the experimental details were provided and the experiment is ready to run. We ask two important questions: what is the size of the ring for which the required small acceleration regime holds and how long is the ring available for a measurement? To answer these questions, we assume that the experiment was prepared so that the axial axes of the ring were such that
%
\begin{equation}
\vec{A} + \vec{A}' \approx (A + A')\hat{u} \lesssim \left(10^{-10}m/s^2\right)\hat{u}.
\label{u}
\end{equation} 
For a ring rotating with angular velocity in the interval (in MKS units)  
%
\begin{equation}
7\times 10^{-2} /\sqrt{R} \lesssim \Omega \lesssim 2\times 10^{-1}/\sqrt{R},
\label{w}
\end{equation}
depending on the latitude where the experiment takes place on Earth's surface, the 
small acceleration regime will be verified only within an angular strip of length 
%
\begin{equation}
\Delta L \lesssim 2R \arcsin{\frac{10^{-10}m/s^2}{A'}}.
\end{equation}
For instance, if the experiment takes place in an equatorial band, the ring's radius being $R=1m$, it can be shown that $\Delta \phi \lesssim 10^{-7} rad$. In other words, the small acceleration regime occurs within in a region of length of the order $\Delta L \lesssim 10^{-7}m$ in the direction of the angular movement of the ring. In orthogonal directions, it occurs in a length given by $2 r a_0/(a_1+a_2) \sim 10^{-1}m$. It means that the small acceleration regime can be measured within an area $\Delta S \lesssim 10^{-8}m^2$ on the ring surface. Additionally, the time interval in which the measurement occurs is of the order $\Delta t \lesssim 10^{-6}s$. Although the experiment must occur in a short time  interval, it can be reloaded several times by preparing the direction of the ring's rotation axes based on the knowledge of $\vec{A}$.

\section{Final remarks} 
The proposed approach to testing the small acceleration regime, by adding a rotating annular device (more generally, a gyroscope), can be carried out at any suitable place on Earth (e.g, the Mauna Kea summit in Hawaii, and Cerro Paranal or Cerro Pach\'on, in Chile, where astronomical observatories are already located), and at any time. This approach is a natural extension of the idea of Ignatiev (2007) of testing fundamental physics at very low acceleration, making his experiment portable to any place on Earth. The price of doing so is a reduced experimental area where the required physical conditions occur, within a shorter interval of time. As the technology of measurement in small areas and short intervals of time increases fast, it is expected that in the near future this type of experiment will be able to be run easily. There are no fundamental restrictions to making these measurements, apart from the high precision of the device involved. At present, the limits stated by the quantum theory are far below the required level of precision. A possible revision of Newton's second law would be required if the results of this experiment indicated that the system exhibited unusual behavior at very low accelerations. In this case, it would be expected that the models already presented in the literature could explain the results of the Milgrom's phenomenological proposal. In terms of astrophysical systems, the possible models that are consistent with the above-mentioned results (if a modification of Newtonian dynamics is required) could be compared directly with cold dark matter theory (Rubin \cite{rubin1983}; Bergstr\"{o}m \cite{berg2000}; Bertone et al. \cite{bert2005}), MONDs proposals (Milgrom \& Sanders \cite{milsan2006}; Sanders \cite{san2006}; Corbelli \& Salucci \cite{corbelli2007}), or with astronomical works such as Famaey \& Binney (\cite{fam2005}), Gentile et al. (\cite{gen2005,gen2008}), Heymans et al. (\cite{hey2006}), and Salucci et al. (\cite{sal2008}).

Several experimental setups could be developed to achieve the regime of very small accelerations, as discussed in this manuscript. In all cases, the main idea is to add an auxiliary device that can produce an extra acceleration such that the total acceleration over a test particle is cancelled out. It is not the scope of this paper to develop the required setup for this in detail. Nevertheless, for the purpose of shedding light on the matter, some details of a possible experiment are presented below: 

\begin{itemize}

\item 
For a $1$-$m$ radius ring, the considerations presented in the 
last section imply an upper limit to the detection area of 
the order of  $10^{-8}m^{2}$. Because of the smallness of the area, 
the width of the ring surface should be large enough to avoid 
possible errors caused by the pointing process (say, a ring surface width of between 1 and 5cm) .

\item 
To measure the effect of small accelerations on test particles, the surface of the ring could consist of a homogeneous crystal lattice of known molecular structure, and mechanical and electronic properties. The characteristics of the crystal, such as the particle diameter, lattice and dielectric constant, Curie temperature, conductance and the lattice exact phonon mode frequencies could be used for the baseline calibration of the crystal lattice (for lattice dynamics theory see Born \& Huang \cite{bh54}; Walmsley \cite{w75}; and Califano et al. \cite{csn81}). In this case, the detection of the small acceleration regime could be achieved statistically by comparing the calibration laboratory spectral pattern (a control template) with the experimental spectral pattern of the frequencies from the reloaded experimental data.

\end{itemize}

The experiment could be improved by decreasing the number of atoms in the crystal lattice (nanocrystallite), thus triggering the decrease in the number of oscillation modes. A nanocrystallite of size 2-3 nanometers consists of N $\approx10^2$ atoms. In this case, the full number of oscillation modes in the lattice of such a nanocrystallite is 3N. With this small number, the oscillation modes are isolated each from other and do not interact among themselves. This provides a possible conceptual design of a detector (see Akulinichev et al. \cite{akk2005} for charged particle detectors development). Within a lattice, all particles (atoms or molecules) are affected by a harmonic potential. For each of these oscillators,  the harmonic potential produces a position-dependent acceleration of  maximum value that is surely beyond the MOND limit. Nevertheless, they all pass by a point of zero acceleration -- the equilibrium point. Thus, there is a region in the neighborhood of this point in which the acceleration due to the harmonic potential is negligible and the considerations discussed in the last section applies. It is expected that the collective effect of  the particle motions in the region of small acceleration produces a pattern that can be compared with the pattern predicted by the Newtonian theory. More specifically, two identical devices (say, $D1$ and $D2$) should be prepared. The axes of one device (the ring axes), $D1$ for instance, may be set in a direction perpendicular to the one defined by acceleration ${\bf A}$, to reach the regime of small acceleration. To ensure that the aforementioned regime is not achieved, $D2$ must be pointed in another direction. In both devices, one must control the matricial time evolution in the velocity-vector map, so that a study of the velocity time-series allows one to extract the power spectrum of the velocity fluctuation on the matricial detecting area of the devices. After measurements have been completed, the spectra obtained from both devices should be compared then. The number of techniques used to analyze signals has steadily grown with time, improving the power of transform-processing algorithms, such as wavelet and Fourier transforms, and synthetic signals, which can be used to analyze the results of both the $D1$ and $D2$ experiments. A non-null result for the subtraction of the spectra would indicate a possible violation of Newton's second law. If this  were the case, the signature of the resulting spectra could be useful in developing a model for the dynamics at low acceleration, and models, such as Milgrom's own could thus be tested using these results.  

On the other hand, knowledge of and technical improvements in measuring the mechanical motions within nanostructures, and laboratory advances in the nanocrystal building, allow one to create functional materials of adjustable structure and properties to be used as direct or indirect detectors (using electronic properties and waves) that could be used to measure small acceleration of test perticles (e.g., the use of the Josephson effect: Josephson \cite{jose62}; Marchenkov et al. \cite{marc07}; and Levy et al. \cite{levy07}; for detectors: Akulinichev et al. \cite{akk2005}; Beck \& Mackey \cite{bm2005} with reply by Jetzer \& Straumann \cite{js05}) and to develop ideas about high-temperature superconductivity in the Josephson granular media (Lebedev \cite{leb2005}).

We provide some concluding remarks to this study. First, as discussed in the last section, the rotating device must produce an acceleration on the test particle in the range given by Eq. (\ref{app}). The exact value depends on the position on Earth where the experiment occurs. Although this acceleration is relatively easy to control, the experiment must be completed so that uncertainties are acceptable only up to the 10th decimal digit. Otherwise the required regime of small accelerations ($\sim a_0$) would not be ``experimentally" achieved. This condition also applies in Ignatiev (\cite{ignatiev2007}), although in this study the difficulty is related to deciding the location for the experiment to be performed.
Second, the precise knowledge of the galactic center of mass can be obtained from the astronomical data. One can indeed identify this center from any place on Earth by means of a position vector defined by the spacetime coordinates of an observer on Earth's surface. We note however that knowledge of this center is necessary only in the case $\vec{a}_2$ is considered in the pointing process. This term is already of order $a_0$ and could well be neglected in the measurement procedure.
Finally, we could also introduce a device with non-fixed axes of rotation (relative to Earth's surface), although, care would then have to be taken to ensure the cancellation of the new contribution to the acceleration produced by the extra degree of freedom.

\begin{acknowledgements}
We wish to thank the anonimous referee for the useful comments that 
helped to improve the presentation of the paper. 
This work was partially supported by the Brazilian funding 
agencies \emph{CNPq\/} and \emph{FAPEMIG\/}.
\end{acknowledgements}

\end{document}